\documentclass[pre,twocolumn,showpacs]{revtex4}
\usepackage{amsmath,amsthm,amssymb,amscd,float}
\usepackage{graphicx}
\usepackage{lmodern}
\newcommand{\al}{\alpha}
\newcommand{\be}{\beta}
\newcommand{\de}{\delta}
\newcommand{\ep}{\epsilon}

\newcommand{\ga}{\gamma}

\newcommand{\si}{\sigma}

\newcommand{\De}{\Delta}

\newcommand{\La}{\Lambda}

\newcommand{\bk}{\mathbf{k}}

\newcommand{\bm}{\mathbf{m}}
\newcommand{\bn}{\mathbf{n}}

\newcommand{\bs}{\mathbf{s}}
\newcommand{\bx}{\mathbf{x}}
\newcommand{\bnu}{{\boldsymbol{\nu}}}

\newcommand{\bxi}{{\boldsymbol{\xi}}}

\newcommand{\tK}{\widetilde{K}}
\newcommand{\tP}{\widetilde{P}}
\newcommand{\tS}{\tilde{S}}

\newcommand{\tih}{\tilde{h}}

\newcommand{\NN}{{\mathbb N}}

\newcommand{\cE}{{\mathcal E}}

\newcommand{\cH}{{\mathcal H}}

\newcommand{\cP}{{\mathcal P}}

\newcommand{\cZ}{{\mathcal Z}}
\def\Esc{E^{\mathrm{sc}}}
\def\Hsc{H^{\mathrm{sc}}}
\def\Zsc{Z^{\mathrm{sc}}}
\newcommand{\pa}{\partial}

\def\ket#1{|#1\rangle}

\let\ni\noindent
\newcommand{\ms}{\mspace{1mu}}
\renewcommand{\le}{\leqslant}
\renewcommand{\ge}{\geqslant}
\renewcommand{\leq}{\leqslant}
\renewcommand{\geq}{\geqslant}
\newcommand{\erf}{\operatorname{erf}}

\newcommand{\tr}{\operatorname{tr}}

\renewcommand{\mod}{\operatorname{mod}}

\newcommand{\e}{\mathrm{e}}

\newcommand{\diff}{\mathrm{d}}
\newcommand{\smax}{s_{\mathrm{max}}}
\newcounter{ex}
\def\cond{\stepcounter{ex}\hskip-.75cm
\makebox[.55cm][r]{(\roman{ex})}\hskip.2cm}
\begin{document}
\title{The Polychronakos--Frahm spin chain of $BC_N$ type and Berry--Tabor's
  conjecture}
\author{J.C. \surname{Barba}}%
\email{jcbarba@fis.ucm.es} \author{F. \surname{Finkel}}%
\email{ffinkel@fis.ucm.es} \author{A. \surname{Gonz\'{a}lez-L\'{o}pez}}
\email[Corresponding author. Electronic address: ]{artemio@fis.ucm.es}
\author{M.A. \surname{Rodr\'{\i}guez}}%
\email{rodrigue@fis.ucm.es} \affiliation{Departamento de F\'{\i}sica
  Te\'{o}rica II, Universidad Complutense, 28040 Madrid, Spain}
\date{March 5, 2008}
\begin{abstract}
We compute the partition function of the su($m$) Polychronakos--Frahm spin chain of $BC_N$ type
by means of the freezing trick. We use this partition function to study several
statistical properties of the spectrum, which turn out to be analogous to those of
other spin chains of Haldane--Shastry type. In particular, we find that when the number of particles
is sufficiently large the level density follows a Gaussian distribution with great accuracy.
We also show that the distribution of (normalized) spacings between consecutive levels is
of neither Poisson nor Wigner type, but is qualitatively similar to that of the
original Haldane--Shastry spin chain. This suggests that spin chains of Haldane--Shastry
type are exceptional integrable models, since they do not satisfy a well-known
conjecture of Berry and Tabor according to which
the spacings distribution of a generic integrable system should be Poissonian.
We derive a simple analytic expression for the cumulative
spacings distribution of the $BC_N$-type Polychronakos--Frahm chain
using only a few essential properties of its spectrum, like the Gaussian character
of the level density and the fact the energy levels are equally spaced. This
expression is in excellent agreement with the numerical data and, moreover, there is strong
evidence that it can also be applied to the Haldane--Shastry and the Polychronakos--Frahm spin chains.
\end{abstract}
\pacs{75.10.Pq, 05.30.-d, 05.45.Mt}
\maketitle
%
\section{Introduction}

Solvable spin chains often provide a natural setting for testing or modeling
interesting physical phenomena and mathematical results in such disparate
fields as fractional statistics, random matrix theory or orthogonal
polynomials. Among these chains, those of Haldane--Shastry (HS) type
occupy a distinguished position due to their remarkable integrability and
solvability properties. The original chain of this type was independently
introduced twenty years ago by Haldane~\cite{Ha88} and
Shastry~\cite{Sh88}, in an attempt to construct a model whose ground state
coincided with Gutzwiller's variational wave function for the Hubbard model in
the limit of large on-site interaction~\cite{Hu63,Gu63,GV87}. In the original
HS chain, the spins are equally-spaced on a circle and present pairwise
interactions inversely proportional to their chord distance.

An essential feature of the spin chains of HS type is their close connection
with the spin versions of the Calogero~\cite{Ca71} and
Sutherland~\cite{Su71,Su72} models, and their generalizations due to
Olshanetsky and Perelomov~\cite{OP83}. This observation ---already pointed out
by Shastry in his original paper--- was elegantly formulated by Polychronakos
in Ref.~\cite{Po93}. In the latter reference, the author showed that the
original HS chain can be obtained from the spin Sutherland
model~\cite{HH92,HW93,MP93} in the strong coupling limit, in which the
dynamical and spin degrees of freedom decouple, so that the particles
``freeze'' at the equilibrium positions of the scalar part of the potential.
In this regime, the integrals of motion of the spin Sutherland model directly
yield first integrals of the HS chain, thereby explaining its complete
integrability. This procedure was also applied in~\cite{Po93} to construct a
new integrable spin chain related to the original Calogero model. The spectrum
of this chain was numerically studied by Frahm~\cite{Fr93}, who found that the
levels are grouped in highly degenerate multiplets. In a subsequent
publication, Polychronakos computed the partition function of this chain
(usually referred to in the literature as the Polychronakos--Frahm chain) by
the ``freezing trick'' argument described above~\cite{Po94}. Interestingly,
the partition function of the original HS chain was computed only very
recently~\cite{FG05}.

Both the HS and the PF (Polychronakos--Frahm) chains are obtained from the
Sutherland and Calogero models associated with the $A_N$ root system in
Olshanetsky and Perelomov's approach. The $BC_N$ versions of both chains have
also been studied in the literature. More precisely, the integrability of the
PF chain of $BC_N$ type was established by Yamamoto and Tsuchiya~\cite{YT96}
using again the freezing trick. On the other hand, the partition function of
the HS chain of $BC_N$ type was computed in closed form in Ref.~\cite{EFGR05}.
The explicit knowledge of the partition function made it possible to study
certain statistical properties of the spectrum of this chain. In particular,
it was observed that for a large number of spins the level density is
Gaussian. As a matter of fact, this property also holds for the original HS
chain, as shown in Ref.~\cite{FG05}. The analysis of the distribution of the
spacing between consecutive levels of the original HS chain was also
undertaken in the latter reference. Rather unexpectedly, it was found that
this distribution is not of Poisson type, as should be the case for a
``generic'' integrable model according to a long-standing conjecture of Berry
and Tabor~\cite{BT77}. This behavior has also been recently reported for a
supersymmetric version of the HS chain~\cite{BB06}.

The aim of this paper is twofold. In the first place, we shall compute in
closed form the partition function of the PF chain of $BC_N$ type by means of
the freezing trick. Using the partition function, we shall perform a numerical
study of the density of levels and the distribution of the spacing between
consecutive energies. We shall see that the level density is again Gaussian,
and that the spacings distribution is analogous to that of the original HS
chain. In particular, our results show that the distribution of spacings is
neither Poissonian nor of Wigner type (characteristic of chaotic systems). We
shall next derive a simple analytic expression for the cumulative spacings distribution,
which reproduces the numerical data with much greater accuracy than the empiric formula
proposed in Ref.~\cite{FG05}. In fact, we have strong numerical evidence that
the new expression can also be applied to the HS and PF chains of $A_N$ type.
In view of the Berry--Tabor conjecture, our results suggest
that spin chains of HS type are exceptional among the class of integrable
models.

\section{The partition function of the PF chain of $BC_N$ type}\label{sec:PF}

The Hamiltonian of the (antiferromagnetic) su($m$) PF chain of $BC_N$ type is
defined by
\begin{equation}\label{cHep}
\cH^\ep=\sum_{i\neq j}\bigg[
\frac{1+S_{ij}}{(\xi_i-\xi_j)^2}+\frac{1+\tS_{ij}}{(\xi_i+\xi_j)^2}\bigg]
+\be\sum_i\frac{1-\ep S_i}{\xi_i^2}\,,
\end{equation}
where the sums run from $1$ to $N$ (as always hereafter, unless otherwise
stated), $\be>0$, $\ep=\pm1$, $S_{ij}$ is the operator which permutes the
$i$-th and $j$-th spins, $S_i$ is the operator reversing the $i$-th spin, and
$\tS_{ij}=S_iS_jS_{ij}$. Note that the spin operators $S_{ij}$ and $S_i$ can
be expressed in terms of the fundamental su($m$) spin generators $J^\al_k$ at
the site $k$ (with the normalization
$\tr(J^\al_kJ^{\ga}_k)=\frac12\de^{\al\ga}$) as
\[
S_{ij}=\frac1m+2\sum_{\al=1}^{m^2-1}J_i^\al J_j^\al\,,\qquad S_i=\sqrt{2m}\,J^1_i\,.
\]
The chain sites $\xi_i$ are the coordinates of the unique minimum in
$C=\{\bx\mid 0<x_1<\cdots<x_N\}$ of the potential
\begin{equation}\label{U}
U(\bx)=\sum_{i\neq j}\bigg[\frac1{(x_{ij}^{-})^2}+\frac1{(x_{ij}^{+})^2}\bigg]
+\sum_i\frac{\be^2}{x_i^2}+\frac{r^2}4\,,
\end{equation}
where $x_{ij}^{\pm}=x_i\pm x_j$ and $r^2=\sum_ix_i^2$. The existence of this minimum
follows from the fact that $U$ tends to $+\infty$ on the
boundary of $C$ and as $r\to\infty$, and its uniqueness
was established in~Ref.~\cite{CS02} by expressing the potential $U$
in terms of the logarithm of the ground state of the $BC_N$ Calogero model
\begin{multline}\label{Hsc}
H^{\mathrm{sc}}=-\sum_i\pa_{x_i}^2+a(a-1)\sum_{i\neq j}\bigg[
\frac1{(x_{ij}^-)^2}+\frac1{(x_{ij}^+)^2}\bigg]\\
+b(b-1)\sum_i\frac1{x_i^2}+\frac{a^2}4\,r^2\,,
\end{multline}
with $b=\be a$ and $a>1/2$. Moreover, it can be shown~\cite{ABCOP79} that
$\xi_i=\sqrt{2y_i}$, where $y_i$ is the $i$-th zero of the generalized Laguerre
polynomial $L_N^{\be-1}$. {}From this fact, one can infer~\cite{CP78c} that
for $N\gg\be$ the density of sites (normalized to unity) $\rho_N(x)$ is given by
the circular law
\begin{equation}
\rho_N(x)=\frac1{2\pi N}\,\sqrt{8N-x^2}\,.
\end{equation}
Note that in this limit the sites' density is independent of $\be$
and is qualitatively similar to that of the PF chain of $A_N$ type~\cite{Fr93}.
Integrating the previous equation, we obtain the implicit asymptotic relation
\[
4\pi k=\xi_k\sqrt{8N-\xi_k^2}+8N\arcsin\Big(\frac{\xi_k}{\sqrt{8N}}\Big)\,,
\]
valid also for $N\gg\be$.

The spin chain~\eqref{cHep} can be expressed in terms of the spin Calogero
model of $BC_N$ type
\begin{multline}\label{Hep}
H^\ep=-\sum_i\pa_{x_i}^2+a\sum_{i\neq j}\bigg[
\frac{a+S_{ij}}{(x_{ij}^-)^2}+\frac{a+\tS_{ij}}{(x_{ij}^+)^2}\bigg]\\
+b\sum_i\frac{b-\ep S_i}{x_i^2}+\frac{a^2}4\,r^2
\end{multline}
and its scalar reduction~\eqref{Hsc} as
\begin{equation}
  \label{Hepchain}
H^\ep=H^{\mathrm{sc}}+a\,\widetilde\cH^\ep\,,
\end{equation}
where $\widetilde\cH^\ep$ is obtained from $\cH^\ep$ replacing the chain sites
$\bxi$ by the particles' coordinates $\bx$. Since
\[
H^\ep=-\sum_i\pa_{x_i}^2+a^2 U+O(a)\,,
\]
when the coupling constant $a$ tends to infinity the particles in the spin
dynamical model~\eqref{Hep} concentrate at the coordinates of the minimum of
the potential $U$, that is at the sites $\xi_i$ of the chain~\eqref{cHep}.
Thus, in the limit $a\to\infty$ the spin and dynamical degrees of freedom of the
Hamiltonian~\eqref{Hep} decouple, so that by Eq.~\eqref{Hepchain} its
eigenvalues are approximately given by
\begin{equation}\label{EEE}
E^\ep_{ij}\simeq E^{\mathrm{sc}}_i+a\cE^\ep_j\,,\qquad a\gg1\,,
\end{equation}
where $E^{\mathrm{sc}}_i$ and $\cE^\ep_j$ are two arbitrary eigenvalues of
$H^{\mathrm{sc}}_i$ and $\cH^\ep$, respectively. The asymptotic
relation~\eqref{EEE} immediately yields the following exact formula for the
partition function $\cZ^\ep$ of the chain~\eqref{cHep}:
\begin{equation}\label{ZZZ}
\cZ^\ep(T)=\lim_{a\to\infty}\frac{Z^\ep(aT)}{Z^{\mathrm{sc}}(aT)}\,,
\end{equation}
where $Z^\ep$ and $Z^{\mathrm{sc}}$ are the partition functions of $H^\ep$ and
$H^{\mathrm{sc}}$, respectively.

We shall next evaluate the partition function $\cZ^\ep$ of the
chain~\eqref{cHep} by computing the partition functions $Z^\ep$ and
$Z^{\mathrm{sc}}$ in Eq.~\eqref{ZZZ}. In order to determine the spectra of the
corresponding Hamiltonians $H^\ep$ and $\Hsc$, following Ref.~\cite{FG05} we
introduce the auxiliary operator
\begin{align}
H'=-\sum_i\pa_{x_i}^2&+\sum_{i\neq j}\bigg[
\frac a{(x_{ij}^-)^2}(a-K_{ij})+\frac a{(x_{ij}^+)^2}(a-\tK_{ij})\bigg]\notag\\
&+\sum_i\frac b{x_i^2}\,(b-K_i)+\frac{a^2}4\,r^2\,,\label{Hp}
\end{align}
where $K_{ij}$ and $K_i$ are coordinate permutation and sign reversing
operators, defined by
\begin{align*}
&(K_{ij}f)(x_1,\dots,x_i,\dots,x_j,\dots,x_N)\\
&\hspace{8.5em}=f(x_1,\dots,x_j,\dots,x_i,\dots,x_N)\,,\\
&(K_i f)(x_1,\dots,x_i,\dots,x_N)=f(x_1,\dots,-x_i,\dots,x_N)\,,
\end{align*}
and $\tK_{ij}=K_iK_jK_{ij}$. We then have the obvious relations
\begin{subequations}
\begin{align}
&H^\ep=H'\big|_{K_{ij}\to-S_{ij},K_i\to\ep S_i}\,,\label{HepHp}\\
&\Hsc=H'\big|_{K_{ij},K_i\to1}\,.\label{HscHp}
\end{align}
\end{subequations}
On the other hand, the spectrum of $H'$ can be easily computed
by noting that this operator can be written in terms of the
rational Dunkl operators of $BC_N$ type~\cite{Du98}
\begin{align}
J_i^-=\pa_{x_i}&+a\sum_{j\neq i}\bigg[\frac1{x_{ij}^-}\,(1-K_{ij})
+\frac1{x_{ij}^+}\,(1-\tK_{ij})\bigg]\notag\\
&+\frac b{x_i}\,(1-K_i)\,,\qquad i=1,\dots,N\,,\label{J-}
\end{align}
as follows~\cite{FGGRZ01b}:
\begin{equation}\label{Hp2}
H'=\mu\Big[-\sum_i\big(J_i^-\big)^2+a\sum_i x_i\pa_{x_i}+E_0\Big]\mu^{-1}\,,
\end{equation}
where
\begin{equation}\label{mu}
\mu(\bx)=\e^{-\frac a4\ms r^2}\prod_i|x_i|^b\cdot\prod_{i<j}{\big|x_i^2-x_j^2\big|}^a
\end{equation}
is the ground state of the Hamiltonian~\eqref{Hsc} and
\begin{equation}\label{E0}
E_0=Na\Big(b+a(N-1)+\frac12\Big)\,.
\end{equation}
Since the Dunkl operators~\eqref{J-} map any monomial $\prod_i x_i^{n_i}$ into a
polynomial of total degree $n_1+\cdots+n_N-1$, by Eq.~\eqref{Hp2} the operator
$H'$ is represented by an upper triangular matrix in the (non-orthonormal)
basis with elements
\begin{equation}
\phi_\bn=\mu\prod_ix_i^{n_i},\:\quad \bn\equiv(n_1,\dots,n_N)\in{(\NN\cup\{0\})}^N,
\end{equation}
ordered according to the total degree $\vert\bn\vert\equiv n_1+\cdots+n_N$ of the
monomial part. More precisely,
\begin{equation}
  \label{Hpphin}
H'\phi_\bn=E'_\bn\phi_\bn+\sum_{\vert\bm\vert<\vert\bn\vert}c_{\bm\bn}\phi_\bm\,,
\end{equation}
where
\begin{equation}\label{Ep}
E'_\bn=a\ms\vert\bn\vert+E_0
\end{equation}
and the coefficients $c_{\bm\bn}$ are real constants.

We shall now construct a basis of the Hilbert space of the Hamiltonian $H^\ep$
in which this operator is also represented by an upper triangular matrix. To
this end, let us denote by $\La^\ep$ the projector on states antisymmetric
under simultaneous permutations of spatial and spin coordinates, and with
parity $\ep$ under sign reversals of coordinates and spins. If
\[
\ket\bs\equiv\ket{s_1,\dots,s_N},\quad s_i=-M,-M+1,\dots,M\equiv{\textstyle\frac{m-1}2},
\]
denotes a state of the su($m$) spin basis, the functions
\begin{equation}\label{psis}
\psi_{\bn,\bs}(\bx)=\La^\ep\big(\phi_\bn(\bx)\ket\bs\big)\,,
\end{equation}
form a basis of the Hilbert space of the Hamiltonian $H^\ep$ provided
that:\smallskip

{\leftskip.75cm\parindent=0pt%
\cond $n_1\geq\cdots\geq n_N$.

\cond $s_i>s_j$ whenever $n_i=n_j$ and $i<j$.

\cond $s_i\geq 0$ for all $i$, and $s_i>0$ if $(-1)^{n_i}=-\ep$.\smallskip

}
\ni The first two conditions are a consequence of the antisymmetry of the
states~\eqref{psis} under particle permutations, while the last condition is
due to the fact that these states must have parity $\ep$ under sign reversals.
Since $K_{ij}\La^\ep=-S_{ij}\La^\ep$ and
$K_i\La^\ep=\ep S_i\La^\ep$, it follows that $H^\ep\La^\ep=H'\La^\ep$. Using
this identity and the fact that $H'$ obviously commutes with $\La^\ep$, from
Eq.~\eqref{Hpphin} we easily obtain
\begin{align*}
H^\ep\psi_{\bn,\bs}&=H'\psi_{\bn,\bs}
=\La^\ep\big((H'\phi_\bn)\ket\bs\big)\\
&=E'_\bn\psi_{\bn,\bs}+\sum_{\vert\bm\vert<\vert\bn\vert}c_{\bm\bn}\psi_{\bm,\bs}\,.
\end{align*}
Thus $H^\ep$ is represented by an upper triangular matrix in the
basis~\eqref{psis}, ordered according to the degree $\vert\bn\vert$. The
diagonal elements of this matrix are given by
\begin{equation}\label{Ens}
E^\ep_{\bn,\bs}=a\ms\vert\bn\vert+E_0\,,
\end{equation}
where $\bn$ and $\bs$ satisfy conditions (i)--(iii) above. Note that, although
the numerical value of $E^\ep_{\bn,\bs}$ is independent of $\bs$, the
degeneracy of each level clearly depends on the spin through the latter
conditions.

Turning next to the scalar Hamiltonian $\Hsc$, in view of Eq.~\eqref{HscHp} we
now need to consider scalar functions of the form
\begin{equation}\label{psissc}
\psi_\bn(\bx)=\La_s\phi_\bn(\bx)\,,
\end{equation}
where $\La_s$ is the symmetrizer with respect to both permutations and sign
reversals. These functions form a (non-orthonormal) basis of the Hilbert space
of $\Hsc$ provided that $n_i=2k_i$ are even integers and $k_1\geq\cdots\geq
k_N$. Just as before, the matrix of the scalar Hamiltonian $\Hsc$ in the
basis~\eqref{psissc} ordered by the degree is upper triangular, with diagonal
elements $\Esc_\bn$ also given by the RHS of~\eqref{Ens}.

Let us next compute the partition functions $\Zsc$ and $Z^\ep$ of the
models~\eqref{Hsc} and~\eqref{Hep}. To begin with, from now on we
shall drop the common ground state energy $E_0$ in both models, since by Eq.~\eqref{ZZZ}
it does not contribute to the partition function $\cZ^\ep$. With this convention,
the partition function of the scalar Hamiltonian $\Hsc$ is given by
\[
\Zsc(aT)=\sum_{k_1\geq\cdots\geq k_N\geq 0}q^{2\vert\bk\vert}\,,
\]
where $q=\e^{-1/(k_{\mathrm B}T)}$. The previous sum can be evaluated by
expressing it in terms of the differences $p_i=k_i-k_{i+1}$, $i=1,\dots,N-1$,
with $p_N\equiv k_N$. Since $k_j=\sum\limits_{i=j}^Np_i$, we easily obtain
\begin{align}
\Zsc(aT)&=\sum_{p_1,\dots,p_N\geq 0}q^{2\sum\limits_{j=1}^N\sum\limits_{i=j}^Np_i}
=\sum_{p_1,\dots,p_N\geq 0}q^{2\sum\limits_{i=1}^Nip_i}\notag\\
&=\prod_i\sum_{p_i\geq 0}{(q^{2i})}^{p_i}=\prod_i(1-q^{2i})^{-1}\,.\label{Zsc}
\end{align}

In order to compute the partition function of the spin Hamiltonian $H^\ep$, we
shall first assume that $m$ is even,
so that condition (iii) simplifies to\smallskip

\ni (iii${}'$) \hskip2mm $s_i>0$ for all $i$.\smallskip

\ni As neither the value of $E^\ep_{\bn,\bs}$ nor conditions (i), (ii) and
(iii${}'$) depend on $\ep$, in this case the partition functions $Z^\ep$ and
$\cZ^\ep$ cannot depend on $\ep$. Hence from now on we shall drop the
superscript $\ep$ when $m$ is even, writing simply $Z$ and
$\cZ$. By Eq.~\eqref{Ens}, after dropping $E_0$ the partition function of the
Hamiltonian~\eqref{Hep} can be written as
\begin{equation}\label{Zep}
Z(aT)=\sum_{n_1\geq\cdots\geq n_N\geq 0}d_\bn\ms q^{\vert\bn\vert}\,,
\end{equation}
where the spin degeneracy factor $d_\bn$ is the number of spin states
$\ket\bs$ satisfying conditions (ii) and (iii${}'$).
Writing
\[
\bn=\big(\overbrace{\vphantom{1}k_1,\dots,k_1}^{\nu_1},\dots,
\overbrace{\vphantom{1}k_r,\dots,k_r}^{\nu_r}\big),\qquad k_1>\cdots>k_r\geq0,
\]
by conditions (ii) and (iii${}'$) we have
\begin{equation}\label{dn}
  d_\bn=\prod\limits_{i=1}^r\binom{m/2}{\nu_i}\equiv d(\bnu)\,,\qquad
  \bnu=(\nu_1,\dots,\nu_r)\,.
\end{equation}
Note that $\sum_{i=1}^r \nu_i=N$, so that the multiindex $\bnu$ can be regarded as an
element of the set $\cP_N$ of partitions of $N$ (taking order into account).
With the previous notation, Eq.~\eqref{Zep} becomes
\begin{align}
  Z(aT)&=\sum_{\bnu\in\cP_N}d(\bnu)
  \sum_{k_1>\cdots>k_r\geq 0}\,q^{\sum\limits_{i=1}^r\nu_i k_i}\notag\\
  &=\sum_{\bnu\in\cP_N}d(\bnu)\sum_{\substack{p_1,\dots,p_{r-1}>0\\p_r\geq 0}}
  q^{\sum\limits_{i=1}^r\nu_i\sum\limits_{j=i}^rp_j}\notag\\
  &=\sum_{\bnu\in\cP_N}d(\bnu)\sum_{\substack{p_1,\dots,p_{r-1}>0\\p_r\geq
      0}}\,
  \prod_{j=1}^rq^{p_j\sum\limits_{i=1}^j\nu_i}\notag\\
  &= q^{-N}\sum_{\bnu\in\cP_N}d(\bnu)\prod_{j=1}^r\frac{q^{N_j}}{1-q^{N_j}}\,,
\label{Zep2}
\end{align}
where
\[
N_j = \sum_{i=1}^j\nu_i\,.
\] {}From Eqs.~\eqref{ZZZ}, \eqref{Zsc} and \eqref{Zep2} we finally obtain the
following explicit expression for the partition function of the
$\mathrm{su}(m)$ PF chain of $BC_N$ type in the case of even $m$:
\begin{equation}
  \label{cZfinal}
  \cZ(T) =
  q^{-N}\prod_i(1-q^{2i})\sum_{\bnu\in\cP_N}d(\bnu)
  \prod_{j=1}^{\ell(\bnu)}\frac{q^{N_j}}{1-q^{N_j}}\,,
\end{equation}
where $\ell(\bnu)=r$ is the number of components of the multiindex $\bnu$.
For instance, for spin $1/2$ we have $\nu_i=1$ for all $i$, and therefore
$r=N$, $d(\bnu)=1$ and $N_j=j$, so that the previous formula simplifies to
\begin{equation}
  \cZ(T) = q^{\frac12N(N-1)}\prod_i(1+q^i)\,,\qquad m=2\,.
  \label{Zschalf}
\end{equation}
Thus, for spin $1/2$ the spectrum of the chain~\eqref{cHep} is given by
\begin{equation}
  \label{spechalf}
\cE_j = \frac12\,N(N-1)+j\,,\qquad j=0,1,\dots, \frac12\,N(N+1)\,,
\end{equation}
and the degeneracy of the energy $\cE_j$ is the number $Q_N(j)$ of partitions
of the integer $j$ into distinct parts no larger than $N$ (with $Q_N(0)\equiv
1$).  For $j\leq N$ this number coincides with the number $Q(j)$ of partitions
of $j$ into distinct parts, which has been extensively studied in the
mathematical literature~\cite{An76}. It is also interesting to observe that
the partition function~\eqref{Zschalf} is closely related to Ramanujan's fifth
order mock theta function~\cite{HAW00}
\[
\psi_1(q)=\sum_{N=0}^\infty q^N\cZ_N(q)\,,
\]
where $\cZ_N(q)$ denotes the RHS of Eq.~\eqref{Zschalf}.

Equation~\eqref{Zschalf} shows that for spin $1/2$ the chain~\eqref{cHep} is
equivalent to a system of $N$ species of noninteracting fermions (with
vacuum energy $\cE_0=N(N-1)/2$), whose effective Hamiltonian is given by
\[
H_{\text{eff}}=\cE_0+\sum_{i=1}^N E_i\,b^\dagger_ib_i\,.
\]
Here $b_i$ (resp.~$b^\dagger_i$) is the annihilation (resp.~creation) operator of
the $i$-th species of fermion, and $E_i=i$ its energy. A similar result was
obtained in Ref.~\cite{BBS08} for the supersymmetric $\mathrm{su}(1|1)$
(ferromagnetic) HS chain, although in the latter case the energy of the $i$-th
fermion is $E_i=i(N-i)$ (the dispersion relation of the original Haldane--Shastry
chain).

Let us consider now the case of odd $m$. In this case, it
is convenient to slightly modify condition (i) above by first grouping the
components of $\bn$ with the same parity and then ordering separately the even
and odd components. In other words, we shall write $\bn=(\bn_{\mathrm
  e},\bn_{\mathrm o})$, where
\begin{align*}
  &\bn_{\mathrm e}=\big(\overbrace{\vphantom{1}2k_1,\dots,2k_1}^{\nu_1},\dots,
  \overbrace{\vphantom{1}2k_s,\dots,2k_s}^{\nu_s}\big),\\
  &\bn_{\mathrm
    o}=\big(\overbrace{\vphantom{1}2k_{s+1}+1,\dots,2k_{s+1}+1}^{\nu_{s+1}},\dots,
  \overbrace{\vphantom{1}2k_r+1,\dots,2k_r+1}^{\nu_r}\big),
\end{align*}
and
\[
k_1>\cdots>k_s\geq0,\qquad k_{s+1}>\cdots>k_r\geq0\,.
\]
By conditions (ii) and (iii), the spin degeneracy factor is now
\begin{equation}\label{dnint}
  d_\bn^\ep=\prod\limits_{i=1}^s\binom{\frac{m+\ep}2}{\nu_i}
  \cdot \prod\limits_{i=s+1}^r\binom{\frac{m-\ep}2}{\nu_i}\equiv d^\ep_s(\bnu)\,.
\end{equation}
Calling
\[
\tilde N_j = \sum_{i=s+1}^j\nu_i\,,\qquad j=s+1,\dots,r\,,
\]
and proceeding as before, we obtain
\begin{align}
  Z^\ep(aT)&=\sum_{\bnu\in\cP_N}\sum_{s=0}^r d^\ep_s(\bnu)
  \hspace*{-10pt}\sum_{\substack{k_1>\cdots>k_s\geq 0\\k_{s+1}>\cdots>k_r\geq
      0}} \hspace*{-10pt}
  q^{\sum\limits_{i=1}^s2\nu_ik_i}q^{\sum\limits_{i=s+1}^r\nu_i(2k_i+1)}\notag\\
  &=\sum_{\bnu\in\cP_N}\sum_{s=0}^r d^\ep_s(\bnu)\,q^{\tilde N_r}\Bigg[
  \sum_{k_1>\cdots>k_s\geq 0}q^{\sum\limits_{i=1}^s2\nu_ik_i}\Bigg]\notag\\
  &\hspace*{6.55em} {}\times\Bigg[\sum_{k_{s+1}>\cdots>k_r\geq 0}
  q^{\sum\limits_{i=s+1}^r2\nu_ik_i}\Bigg]\notag\\
  &=\sum_{\bnu\in\cP_N}\sum_{s=0}^{\ell(\bnu)}d^\ep_s(\bnu)\,q^{-(N+N_s)}
  \prod_{j=1}^s\frac{q^{2N_j}}{1-q^{2N_j}}\notag\\
  &\hspace*{6.55em} {}\times\prod_{j=s+1}^{\ell(\bnu)}\frac{q^{2\tilde
      N_j}}{1-q^{2\tilde N_j}}\,.
\end{align}
Substituting the previous expression and~\eqref{Zsc} into \eqref{ZZZ}, we
immediately deduce the following explicit formula for the partition functions
of the $\mathrm{su}(m)$ PF chain of $BC_N$ type for odd $m$:
\begin{multline}
  \label{cZodd}
  \cZ^\ep(T)=\prod_i(1-q^{2i})\sum_{\bnu\in\cP_N}
  \sum_{s=0}^{\ell(\bnu)}d^\ep_s(\bnu)\,q^{-(N+N_s)}\\
  {}\times\prod_{j=1}^s\frac{q^{2N_j}}{1-q^{2N_j}}\,\,\cdot\!\!
  \prod_{j=s+1}^{\ell(\bnu)}\frac{q^{2\tilde N_j}}{1-q^{2\tilde N_j}}\,.
\end{multline}

Although we have chosen, for definiteness, to study the antiferromagnetic
chain~\eqref{cHep}, a similar analysis can be performed for its ferromagnetic
counterpart
\begin{equation}\label{cHepF}
\cH^\ep_{\mathrm F}=\sum_{i\neq j}\bigg[
\frac{1-S_{ij}}{(\xi_i-\xi_j)^2}+\frac{1-\tS_{ij}}{(\xi_i+\xi_j)^2}\bigg]
+\be\sum_i\frac{1-\ep S_i}{\xi_i^2}\,.
\end{equation}
Since now
\begin{equation}
\label{HepFHp}
H^\ep_{\mathrm F}=H'\big|_{K_{ij}\to S_{ij},K_i\to\ep S_i}\,,
\end{equation}
we must replace the operator $\Lambda^\ep$ in Eq.~\eqref{psis} by the projector on
states \emph{symmetric} under simultaneous permutations of the particles'
spatial and spin coordinates, and with parity $\ep$ under sign reversal of
coordinates and spin. Hence condition (ii) above on the basis
states~$\psi_{\bn,\bs}$ should now read

\smallskip\noindent
(ii${}'$) $s_i\geq s_j$ whenever $n_i=n_j$ and $i<j$.

\smallskip\noindent As a result, the degeneracy
factors $d(\bnu)$ and $d^\ep(\bnu)$ in Eqs.~\eqref{dn} and~\eqref{dnint}
should be replaced by their ``bosonic'' versions
\begin{align*}
d_{\mathrm F}(\bnu)&=\prod\limits_{i=1}^r\binom{\frac m2+\nu_i-1}{\nu_i}\,,\\
d_{\mathrm F}^\ep(\bnu)&=\prod\limits_{i=1}^s\binom{\frac{m+\ep}2
+\nu_i-1}{\nu_i}
\cdot \prod\limits_{i=s+1}^r\binom{\frac{m-\ep}2+\nu_i-1}{\nu_i}\,.
\end{align*}
Therefore the partition function of the ferromagnetic $\mathrm{su}(m)$ PF
chain of $BC_N$ type~\eqref{cHepF} is still given by Eq.~\eqref{cZfinal} (for
even $m$) or~\eqref{cZodd} (for odd $m$), but with $d(\bnu)$ and $d^\ep(\bnu)$
replaced respectively by $d_{\mathrm F}(\bnu)$ and $d^\ep_{\mathrm F}(\bnu)$.

On the other hand, the chains~\eqref{cHep} and~\eqref{cHepF} are obviously
related by
\[
\cH^\ep_{\mathrm F}+\cH^{-\ep}=2\Big(\sum_{i\neq
  j}\big[(\xi_i-\xi_j)^{-2}+(\xi_i+\xi_j)^{-2}\big]
+\be\sum_i\xi_i^{-2}\Big)\,.
\]
The RHS of this equation clearly coincides with the largest eigenvalue
$\cE_{\mathrm{max}}$ of the antiferromagnetic chains $\cH^\ep$, whose
corresponding eigenvectors are the spin states symmetric under permutations
and with parity $\ep$ under spin reversal. This eigenvalue is most easily
computed for the spin $1/2$ chains, since in this case the spectrum is
explicitly given in~Eq.~\eqref{spechalf}. We thus obtain
\begin{equation}\label{cEmax}
\cE_{\mathrm{max}}=\frac12\,N(N-1)+\frac12\,N(N+1)=N^2\,,
\end{equation}
so that
\[
\cH^\ep_{\mathrm F}=N^2-\cH^{-\ep}\,.
\]
Hence the partition functions $\cZ^\ep$ and $\cZ_{\mathrm F}^\ep$ of $\cH^\ep$
and $\cH^\ep_{\mathrm F}$ satisfy the remarkable identity
\[
\cZ_{\mathrm F}^\ep(q)=q^{N^2}\cZ^{-\ep}(q^{-1})\,.
\]
This is a manifestation of the boson-fermion duality discussed in detail in
Ref.~\cite{BBHS07} for the $\mathrm{su}(m|n)$ supersymmetric HS spin chain,
since the ferromagnetic (resp.~antiferromagnetic) chain can be regarded as
purely bosonic (resp.~fermionic). For instance, using the latter identity and
Eq.~\eqref{Zschalf} we easily obtain the following expression for the
partition function of the ferromagnetic spin $1/2$ chains:
\begin{equation}\label{cZF2}
\cZ_{\mathrm{F}}(T)=\prod_i(1+q^i)\,,\qquad m=2\,.
\end{equation}
(Note that, as in the antiferromagnetic case, $\cZ^\ep_{\mathrm{F}}$ is
actually independent of $\ep$ for even $m$.) This is, again, the partition
function of a system of $N$ species of free fermions of energy $E_i=i$, but
now the vacuum energy vanishes.

Equation~\eqref{cZfinal} for the partition function of the
antiferromagnetic chains with even $m$ can be easily simplified to
\[
\cZ(T) = \prod_i(1+q^{i})\sum_{\bnu\in\cP_N}d(\bnu)\,
q^{\sum\limits_{j=1}^{\ell(\bnu)-1}\kern-5pt N_j}\,
\prod_{j=1}^{N-\ell(\bnu)}(1-q^{N_j'})\,,
\]
where the positive integers $N'_j$ are defined by
\[
\big\{N_1',\dots,N'_{N-\ell(\bnu)}\big\} = \big\{1,\dots,N-1\big\}
-\big\{N_1,\dots,N_{\ell(\bnu)-1}\big\}.
\]
The sum in the RHS is easily recognized as the partition function
$\cZ^{(A)}(T;m/2)$ of the $\mathrm{su}(m/2)$ (antiferromagnetic) PF chain of
$A_N$ type~\cite{BB06}. We thus obtain the remarkable factorization
\begin{equation}
  \label{cZprod}
  \cZ(T;m) = \cZ_{\mathrm{F}}(T;2)\cdot\cZ^{(A)}(T;m/2)\,,\quad
  m\in2\NN\,,
\end{equation}
where the second argument in $\cZ$ and $\cZ_{\mathrm F}$ denotes
the number of internal degrees of freedom. Replacing $d(\bnu)$ by
$d_{\mathrm{F}}(\bnu)$ in Eq.~\eqref{cZfinal} we obtain a similar
factorization for the partition function of the ferromagnetic chains:
\begin{equation}
  \label{cZprodF}
  \cZ_{\mathrm F}(T;m) = \cZ_{\mathrm{F}}(T;2)\cdot\cZ^{(A)}_{\mathrm
  F}(T;m/2)\,,\quad
  m\in2\NN\,.
\end{equation}
Thus, for even $m$ the PF chains of $BC_N$ type~\eqref{cHep} and \eqref{cHepF}
can be described by an effective model of two simpler noninteracting chains.
This remarkable property, which to the best of our knowledge is unique among
the class of chains of Haldane--Shastry type, certainly deserves further
investigation.

\section{Spacings distribution and the Berry--Tabor conjecture}

For fixed values of the number of particles $N$ and the internal degrees of
freedom $m$, it is straightforward to obtain the spectrum of the
chain~\eqref{cHep} by expanding in powers of $q$ the expressions~\eqref{cZfinal} or \eqref{cZodd}
for its partition function. In this way, we have been able to
compute the spectrum of the latter chain for relatively large values of $N$ (for instance, up
to $N=22$ for $m=3$). Our calculations conclusively show that the spectrum consists of a set
of consecutive integers. For even $m$, this observation
follows immediately from the expression~\eqref{cZF2}-\eqref{cZprod} and the fact the
the energies of the PF chain of $A_N$ type are also consecutive integers.
For odd $m$ we have been unable to deduce this property from Eq.~\eqref{cZodd} for the partition
function, although we have verified it numerically for many different values of $N$ and $m$.

\begin{figure}[h]
	\centering
        \includegraphics[width=8.6cm]{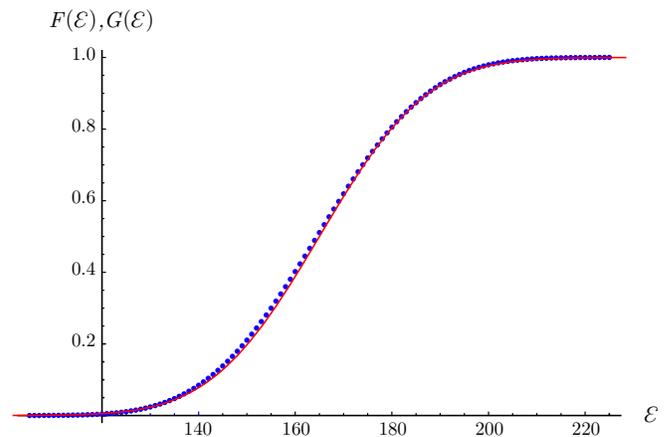}
                \caption{Cumulative distribution functions $F(\cE)$ (at its
                  discontinuity points) and $G(\cE)$ (continuous red line) for
                  $N=15$ and spin $1/2$.}
	\label{fig:levden}
\end{figure}

Our computations also evidence that for $N\gtrsim 10$ the
level density (normalized to unity) can be approximated with great accuracy by a
normal distribution
\begin{equation}\label{Gaussian}
g(\cE)=\frac{1}{\sqrt{2\pi}\si}\,\e^{-\frac{(\cE-\mu)^2}{2\si^2}}\,,
\end{equation}
where $\mu$ and $\si$ are respectively the mean and the variance of the energy.
For instance, in Fig.~\ref{fig:levden} we compare the cumulative level density
\[
F(\cE)=m^{-N}\sum_{i;\cE_i\leq \cE}d_i\,,
\]
where $\cE_i$ is the $i$-th energy and $d_i$ its degeneracy, with the
cumulative Gaussian density
\begin{equation}\label{G}
G(\cE)=\int_{-\infty}^{\cE} g(\cE')\,\diff\cE' =
\frac12\bigg[1+\erf\bigg(\frac{\cE-\mu}{\sqrt 2\,\si}\bigg)\bigg]
\end{equation}
for $N=15$ and $m=2$.
Note, in this respect, that the approximately Gaussian character of the level
density has already been verified for other chains of HS type, like the
original Haldane--Shastry chain~\cite{FG05}, its $\mathrm{su}(m|n)$
supersymmetric version~\cite{BB06}, and the HS spin chain of $BC_N$
type~\cite{EFGR05}.

The mean energy $\mu$ and its standard deviation $\sigma^2$, which by the
previous discussion characterize the approximate level density of the
chain~\eqref{cHep} for large $N$, can be computed in closed form. Indeed, in
Appendix~\ref{app:musigma} we show that
\begin{align}
  \label{mean}
  \mu&=\frac12\,\Big(1+\frac1m\Big)N^2-
  \frac N{2m}\,(1+\ep\ms p),\\
  \si^2&=\frac N{36}\,(4 N^2+6N-1)\Big(1-\frac1{m^2}\Big)+\frac{N(1-p)}{4m^2}\,,
  \label{sigma2}
\end{align}
where $p\in\{0,1\}$ is the parity of $m$. Thus, when $N$ tends to infinity $\mu$
and $\si^2$ respectively diverge as $N^2$ and $N^3$, as for the original
Polychronakos--Frahm chain~\footnote{Work in progress by the authors.}.
By contrast, it is known that $\mu\sim N^3$ and $\si^2\sim N^5$ for
the trigonometric HS chains of both $A_N$~\cite{FG05} and $BC_N$~\cite{EFGR05} types.
It is also interesting to observe that the standard deviation of the energy is
independent of $\ep$ even for odd $m$, when the spectrum does depend on $\ep$
according to the previous section's results on the partition function.

We have next studied the probability density $p(s)$ of the spacing $s$ between consecutive
(unfolded) levels of the chain~\eqref{cHep}. For many important integrable
systems it is known that $p(s)$ is Poissonian~\cite{PZBMM93,AMV02}, in
agreement with a well-known conjecture of Berry and Tabor~\cite{BT77}. On the
other hand, it has been recently shown that for the HS chain of $A_N$
type~\cite{FG05} (and its supersymmetric extension~\cite{BB06}) the cumulative
density $P(s)=\int_0^s p(x)\diff x$ is well approximated by an empiric law of
the form
\begin{equation}\label{empP}
\tP(s)=\bigg(\frac s{s_{\mathrm{max}}}\bigg)^\al\Bigg[1-\ga\bigg(1-\frac s{s_{\mathrm{max}}}\bigg)^\be\Bigg]\,,
\end{equation}
where $s_{\mathrm{max}}$ is the largest normalized spacing and $\al,\be$ are
adjustable parameters in the interval $(0,1)$. The parameter $\ga$ is fixed by
requiring that the average spacing be equal to~$1$, with the result
\begin{equation}\label{ga}
\ga=\Big(\frac 1{s_{\mathrm{max}}}-\frac\al{\al+1}\Big)\Big/B(\al+1,\be+1)\,,
\end{equation}
where $B$ is Euler's Beta function. Thus, the cumulative density of spacings for the HS chains
of $A_N$ type follows neither Poisson's nor Wigner's
law
\[
P(s)=1-\e^{-\pi s^2/4},
\]
characteristic of a chaotic system. Our aim is to ascertain whether the cumulative density of spacings for
the PF chain of $BC_N$ type~\eqref{cHep} resembles that of the $A_N$-type HS chain, or is rather
Poissonian as expected for a generic integrable model.

In order to compare the spacings distributions of spectra with
different level densities, it is necessary to transform the ``raw'' spectrum
by applying what is known as the \emph{unfolding} mapping~\cite{Ha01}. This
mapping is defined by decomposing the cumulative level density $F(\cE)$ as the
sum of a fluctuating part $F_{\mathrm{f{}l}}(\cE)$ and a continuous part
$\eta(\cE)$, which is then used to transform each energy $\cE_i$,
$i=1,\dots,n$, into an unfolded energy $\eta_i=\eta(\cE_i)$. In this way one
obtains a uniformly distributed spectrum $\{\eta_i\}_{i=1}^n$, regardless of
the initial level density. One finally considers the normalized spacings
$s_i=(\eta_{i+1}-\eta_i)/\De$, where $\De=(\eta_{n}-\eta_1)/(n-1)$ is the mean
spacing of the unfolded energies, so that $\{s_i\}_{i=1}^{n-1}$ has unit mean.

\begin{figure}[h]
\centering
\includegraphics[width=8.6cm]{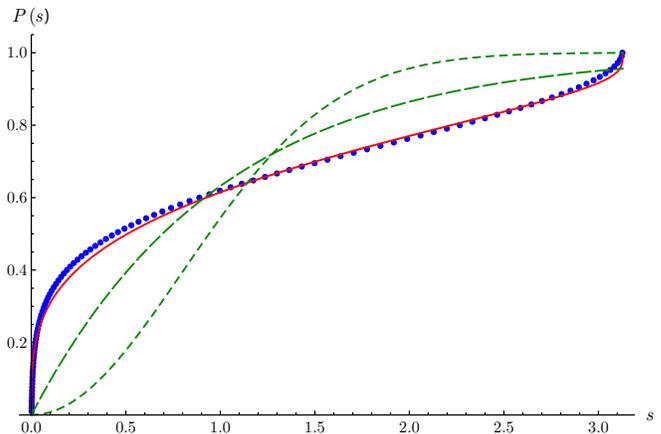}
\caption{Cumulative spacings distribution $P(s)$ and its approximation
  $\tP(s)$ (continuous red line) for $N=20$ and $m=2$. For convenience, we
  have also represented Poisson's (green, long dashes) and Wigner's (green,
  short dashes) cumulative distributions.\label{fig:accbeta}}
\end{figure}

By the above discussion, in our case we can take the unfolding mapping $\eta(\cE)$
as the cumulative Gaussian distribution~\eqref{G}, with parameters
$\mu$ and $\si$ respectively given by~\eqref{mean} and~\eqref{sigma2}. Just as for
the level density, in order to compare the discrete distribution function
$p(s)$ with a continuous distribution it is more convenient to
work with the cumulative spacings distribution $P(s)$. Our computations
show that for a wide range of values of $N$, $m$ and $\ep=\pm1$,
the distribution $P(s)$ is well approximated by the empiric law~\eqref{empP}
with suitable values of $\al$ and $\be$. For instance, for $N=20$ and $m=2$
the largest spacing is $s_{\mathrm{max}}=3.13$, and the least-squares fit
parameters $\al$ and $\be$ are respectively $0.29$ and $0.24$, with
a mean square error of $6.0{{\times}} 10^{-4}$ (see Fig.~\ref{fig:accbeta}).
Thus the PF spin chain of $BC_N$ type behaves in this respect as the HS chain
of $A_N$ type, and unlike most known integrable systems. In fact, we have also
studied the spacings' distribution of the original ($A_N$-type) PF chain,
obtaining completely similar results~[35]. These results (and also those of
Ref.~\cite{BB06}) suggest that a spacings distribution qualitatively similar
to the empiric law~\eqref{empP} is characteristic of all spin chains of HS
type.

Our next objective is to explain this characteristic behavior of the cumulative spacings distribution $P(s)$
of the chain~\eqref{cHep} using only a few essential properties of its spectrum. We shall find a
simple analytic expression without any adjustable parameters approximating $P(s)$ even better
than the empiric law~\eqref{empP}. Moreover, we have strong numerical evidence
that the new expression also provides a very accurate approximation to the cumulative spacings
distribution of the original HS and PF chains.

Consider, to begin with, any spectrum $\cE_1\equiv\cE_{\mathrm{min}}<\cdots<\cE_n\equiv\cE_{\mathrm{max}}$
obeying the following conditions:

{\leftskip.75cm\parindent=0pt\setcounter{ex}{0}%
\cond The energies are equispaced, i.e., $\cE_{i+1}-\cE_i=\de$ for $i=1,\dots,n-1$.

\cond The level density (normalized to unity) is approximately Gaussian, cf.~Eq.~\eqref{Gaussian}.

\cond $\cE_n-\mu\,,\,\mu-\cE_1\gg\si$.

\cond $\cE_1$ and $\cE_n$ are approximately symmetric with respect to $\mu$, namely
$\vert\cE_1+\cE_n-2\mu\vert\ll\cE_n-\cE_1$.\smallskip

}
\ni As discussed above, the spectrum of the chain~\eqref{cHep} satisfies the first
condition with $\de=1$, while condition (ii) holds for sufficiently large $N$. As to
the third condition, from Eqs.~\eqref{cEmax}, \eqref{mean}, \eqref{sigma2},
\eqref{cEmineven} and~\eqref{cEminodd} it follows that both
$(\cE_n-\mu)/\si$ and $(\mu-\cE_1)/\si$ grow as $N^{1/2}$ when $N\to\infty$.
The last condition is also satisfied for large $N$, since by the equations
just quoted $\vert\cE_1+\cE_n-2\mu\vert=O(N)$ while $\cE_n-\cE_1=O(N^2)$.

\begin{figure}[h]
\centering
\includegraphics[width=8.6cm]{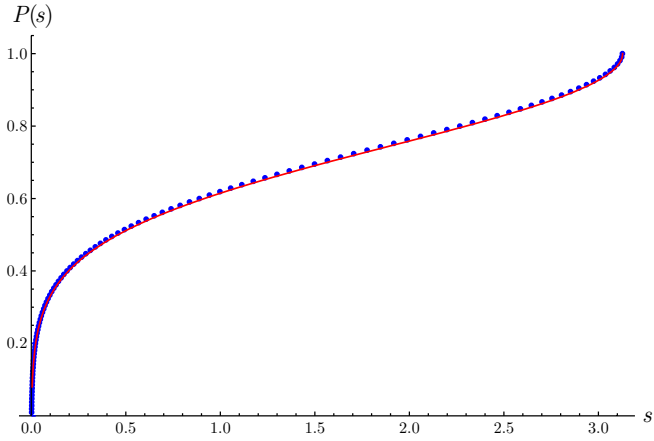}
\caption{Cumulative spacings distribution $P(s)$ and its analytic approximation~\eqref{P}
(continuous red line) for $N=20$ and $m=2$.\label{fig:P}}
\end{figure}

{}From conditions (i) and (ii) it follows that
\[
\eta_{i+1}-\eta_i=G(\cE_{i+1})-G(\cE_i)\simeq G'(\cE_i)\de=g(\cE_i)\de\,.
\]
On the other hand, by condition (iii) we have
\[
\eta_n=G(\cE_n)\simeq 1\,,\qquad \eta_1=G(\cE_1)\simeq 0\,,
\]
so that $\De=1/(n-1)$. Thus
\begin{equation}\label{si}
s_i=\frac{\eta_{i+1}-\eta_i}\De\simeq Wg(\cE_i)=
\frac{W}{\sqrt{2\pi}\si}\,\e^{-\frac{(\cE_i-\mu)^2}{2\si^2}}\,,
\end{equation}
where
\begin{equation}\label{W}
W\equiv(n-1)\de=\cE_n-\cE_1
\end{equation}
on account of the first condition. The cumulative probability density $P(s)$ is
by definition the quotient of the number of normalized spacings $s_i\le s$
by the total number of spacings, that is,
\[
P(s)=\frac{\#(s_i\le s)}{n-1}\,.
\]
By Eq.~\eqref{si},
\begin{equation}
  \label{nsi}
  \#(s_i\le s)=
  \#\big(\cE_1\le\cE_i\le\cE_-\big)+\#\big(\cE_+\le\cE_i<\cE_n\big)\,,
\end{equation}
where
\begin{equation}\label{cEpm}
\cE_\pm=\mu\pm\sqrt2\,\si\sqrt{\log\Big(\frac{\smax}s\Big)}
\end{equation}
are the roots of the equation $s=W g(\cE)$ expressed in terms of the maximum normalized spacing
\begin{equation}\label{smax}
\smax=\frac W{\sqrt{2\pi}\si}\,.
\end{equation}
Using the first condition to estimate the RHS of Eq.~\eqref{nsi} we easily
obtain
\begin{equation}
  \label{Pmax}
  P(s)\simeq\frac1W\,\big[\max(\cE_--\cE_1,0)+\max(\cE_n-\cE_+,0)\big]\,.
\end{equation}
In fact, we can replace the latter approximation to $P(s)$ by the simpler one
\begin{equation}
  \label{Papprox}
    P(s)\simeq\frac1W\,\big(\cE_--\cE_1+\cE_n-\cE_+\big)\,,
\end{equation}
since the error involved is bounded by
\[
\frac1W\,\big|\cE_1+\cE_n-2\mu\big|\,,
\]
which is vanishingly small by condition (iv). Substituting the explicit expression~\eqref{cEpm}
for $\cE_\pm$ into Eq.~\eqref{Papprox} and using~\eqref{W} and~\eqref{smax} we finally obtain
\begin{equation}
  \label{P}
  P(s)\simeq 1-\frac{2}{\sqrt\pi\,\smax}\,\sqrt{\log\Big(\frac{\smax}s\Big)}\,.
\end{equation}
The RHS of this remarkable expression depends only on the quantity $\smax$, which for the PF chain of $BC_N$ type
is completely determined as a function of $N$ and $m$ by Eqs.~\eqref{cEmax},  \eqref{W}, \eqref{smax} and
\eqref{cEmineven}-\eqref{cEminodd}. In particular, for large $N$ we have the asymptotic expression
\[
\smax=\frac3{\sqrt{2\pi}}\,\sqrt{\frac{m-1}{m+1}}\:N^{1/2}+O\big(N^{-1/2}\big)\,.
\]
Our numerical computations indicate that Eq.~\eqref{P} is in excellent agreement with the data
for a broad range of values of $N$, $m$ and $\ep=\pm1$,
providing much greater accuracy than the empiric formula~\eqref{empP}.
For instance, for $N=20$ and $m=2$ we have $\smax=6\sqrt{35/(41\pi)}\simeq 3.12765$,
which differs from the numerically computed maximum spacing by $4.5\times 10^{-4}$.
In Fig.~\ref{fig:P} we compare the corresponding cumulative spacings distribution
with its approximation~\eqref{P} using the above value of $\smax$. The mean square
error is in this case $2.2\times 10^{-5}$, smaller than that of the empiric law~\eqref{empP}
by more than an order of magnitude.

\begin{figure}[h]
\centering
\includegraphics[width=4.2cm]{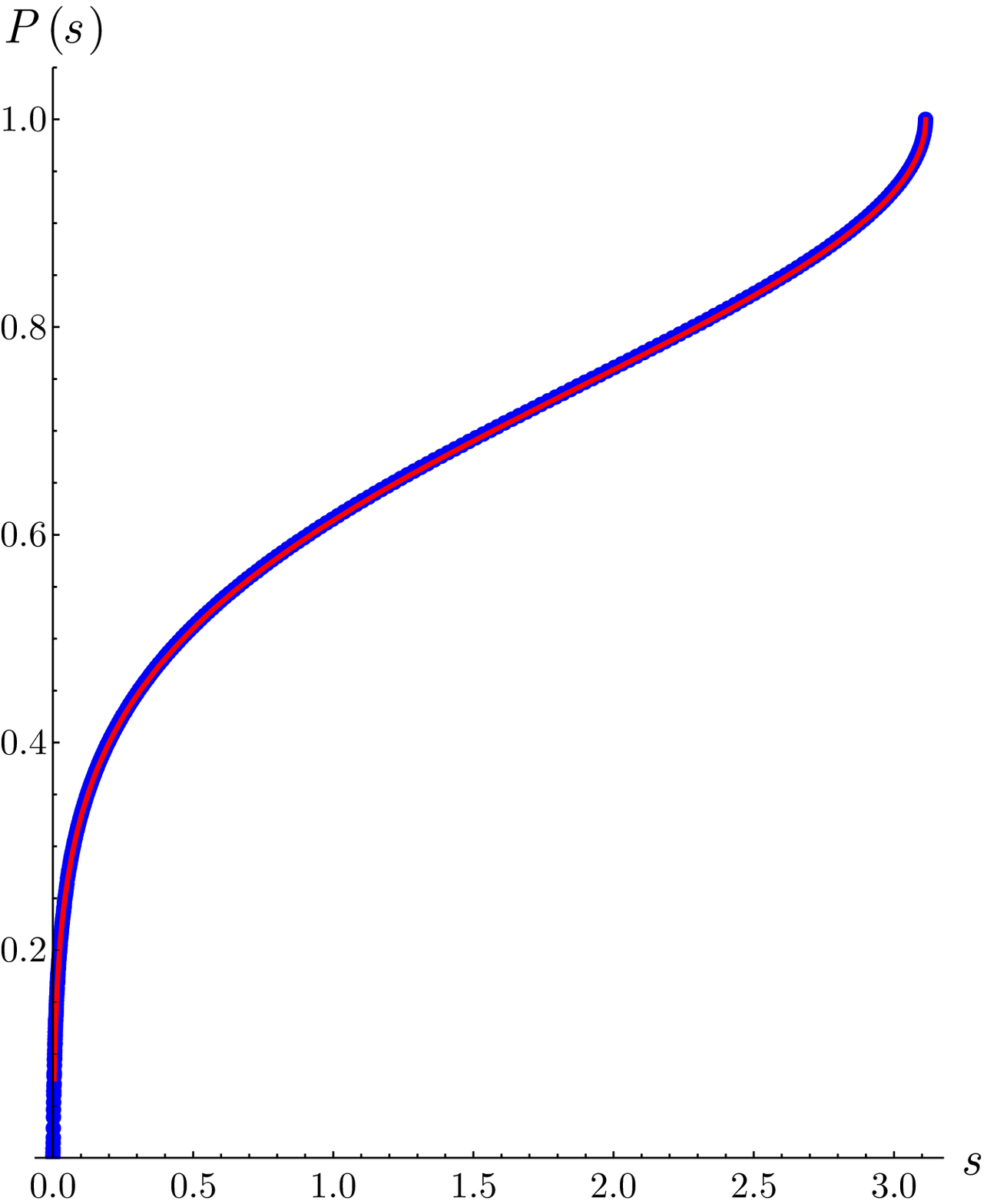}\hfill
\includegraphics[width=4.2cm]{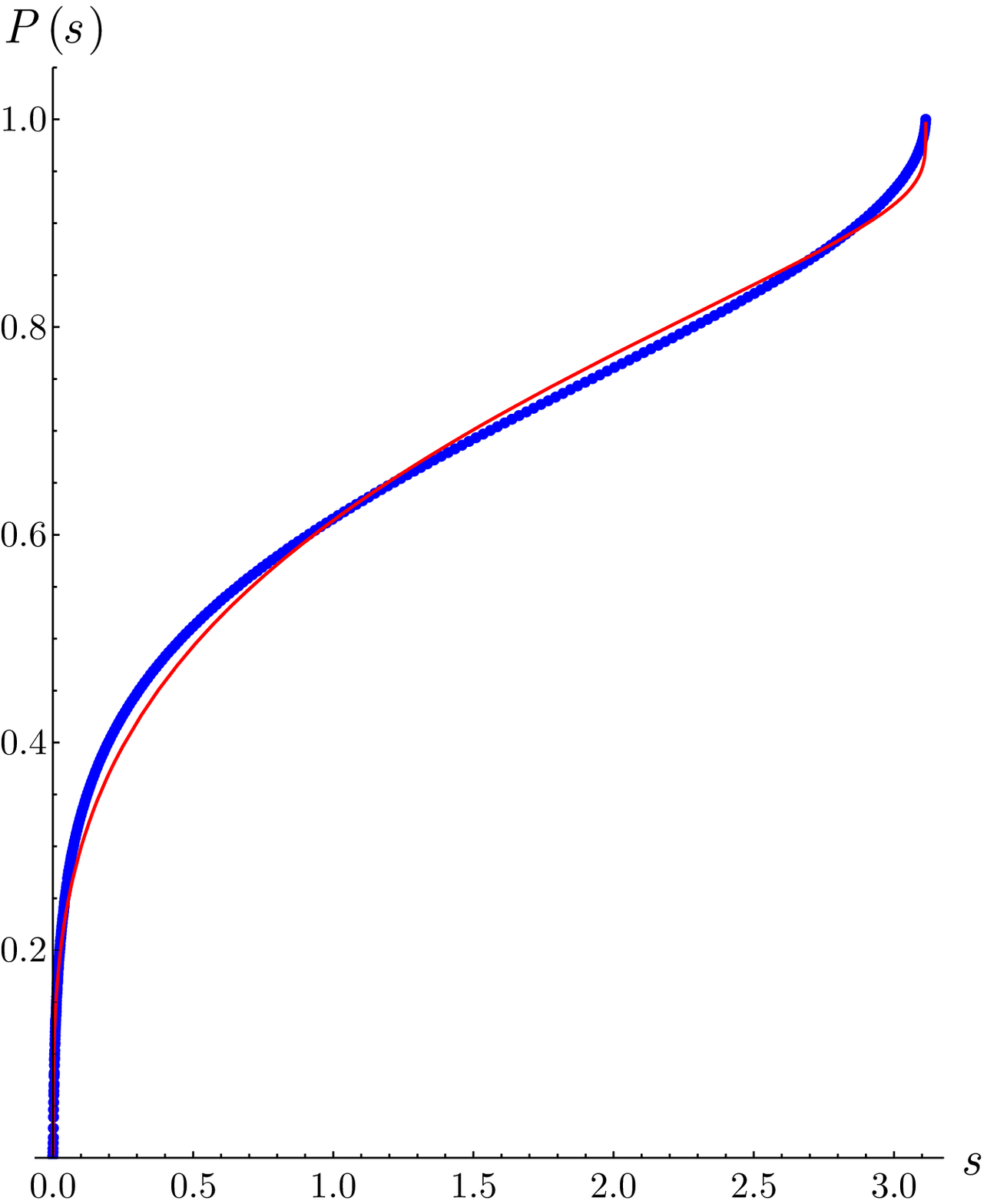}
\caption{Cumulative spacings distribution $P(s)$ for the HS chain of $A_N$ type
in the case of $N=25$ and $m=2$, compared to its continuous approximations~\eqref{P} (left)
and~\eqref{empP} (right). In both cases, the thin (red) line corresponds to the
continuous approximation.\label{fig:PHSA}}
\end{figure}

A natural question in view of these results is to what extent the
approximation~\eqref{P} is applicable to other spin chains of HS type. For the
PF chain of $A_N$ type, one can check that the spectrum satisfies conditions
(i)--(iv) of this section, and in fact we have verified that~\eqref{P} holds
with remarkable accuracy in this case~[35]. The situation is less clear for
the original HS chain, whose spectrum is certainly not equispaced~\cite{FG05}.
Nevertheless, our computations show that the formula~\eqref{P} still fits the
numerical data much better than our previous approximation~\eqref{empP}, a
fact clearly deserving further study. As an illustration, in
Fig.~\ref{fig:PHSA} we compare the cumulative spacings distribution of the
original HS chain with its approximations~\eqref{empP} and~\eqref{P} for
$N=25$ and $m=2$. It is apparent that the new expression~\eqref{P} provides a
more accurate approximation to the numerical data than the empiric
formula~\eqref{empP} (their respective mean square errors are $1.8\times 10^{-5}$
and $5.8\times 10^{-4}$).

\appendix

\section{Computation of $\mu$ and $\sigma^2$}
\label{app:musigma}

In this appendix we shall compute in closed form the mean energy $\mu$ of the
spin chain~\eqref{cHep} and its standard deviation $\sigma^2$ as functions of
the number of particles $N$ and internal degrees of freedom $m$.

In the first place, using the formulas for the traces of the operators
$S_{ij}$, $\tS_{ij}$ and $S_i$ in Ref.~\cite{EFGR05} we easily obtain
\begin{align}
  \mu &= \frac{\tr\cH_\ep}{m^N}\notag\\
  \label{muexp}
  &= \Big(1+\frac1m\Big)\sum_{i\neq
    j}(h_{ij}+\tih_{ij})+\Big(1-\frac{\ep\ms p}m\Big)\sum_i h_i\,,
\end{align}
where $p$ is the parity of $m$ and
\[
h_{ij}=(\xi_i-\xi_j)^{-2}\,,\quad \tih_{ij}=(\xi_i+\xi_j)^{-2}\,,\quad
h_i=\be\,\xi_i^{-2}\,.
\]
The sums appearing in Eq.~\eqref{muexp} can be expressed in terms of sums
involving the zeros $y_i=\xi_i^2/2$, $1\leq i\leq N$, of the Laguerre polynomial
$L_N^{\be-1}$ as follows:
\begin{equation}
  \label{hijhi}
    \sum_{i\neq j}(h_{ij}+\tih_{ij})
    =\sum_{i\neq j}\frac{2y_i}{(y_i-y_j)^2}\,,\quad
    \sum_ih_i = \frac\be2\sum_i\frac1{y_i}\,.
\end{equation}
The latter sums can be easily computed using the following identities
satisfied by the zeros $y_i$, which
can be found in Ref.~\cite{Ah78}:
\begin{align}
  \label{sum1}
  &\sum_{j,j\neq i}\frac{2\,y_i}{y_i-y_j}=y_i-\be\,,\\
  \label{sum2}
  &\sum_{j,j\neq i}\frac{12\,y_i^2}{(y_i-y_j)^2}=-y_i^2+2(2N+\be)y_i-\be(\be+4)\,.
\end{align}
Indeed, from the first of these identities we easily obtain
\begin{equation}
\sum_iy_i =N(N+\be-1)\,,\qquad
\sum_i\frac1{y_i}=\frac N\be\,,
\label{yiyim1}
\end{equation}
so that, by Eq.~\eqref{sum2},
\begin{multline*}
\sum_{i\neq j}\frac{2\,y_i}{(y_i-y_j)^2}=\frac16\,\Big(-\sum_i
y_i+2N(2N+\be)\\-\be(\be+4)\sum_i\frac1{y_i}\Big)
= \frac12\,N(N-1)\,.
\end{multline*}
Combining the last two equations with~\eqref{muexp} and~\eqref{hijhi} we
immediately arrive at Eq.~\eqref{mean} for the level density $\mu$.

Turning now to the standard deviation of the energy $\si^2$,
from Eqs.~(66)--(68) in Ref.~\cite{EFGR05} we have
\begin{multline}
  \label{sigma2exp}
  \si^2= \frac{\tr\cH_\ep^2}{m^N}-\mu^2 =\Big(1-\frac1{m^2}\Big)
  \Big(2\sum_{i\neq j}\big(h_{ij}^2+\tih_{ij}^2\big)+\sum_i h_i^2\Big)\\
  +\frac4{m^2}\,(1-p)\Big(\frac14\sum_i h_i^2-\sum_{i\neq
    j}h_{ij}\tih_{ij}\Big)\,.
\end{multline}
All of the sums appearing in the latter expression can be readily evaluated.
Indeed, we have
\begin{align}
  \label{hi2}
  &\sum_i h_i^2=\frac{\be^2}4\sum_i\frac1{y_i^2}=\frac{N(N+\be)}{4(\be+1)}\\
  \label{hijthij}
  &\sum_{i\neq j}h_{ij}\tih_{ij}=\frac14\sum_{i\neq
    j}\frac1{(y_i-y_j)^2}=\frac{N(N-1)}{16(\be+1)}\,,
\end{align}
where we have used Eqs.~(15) and (17) from Ref.~\cite{Ah78}. On the other
hand,
\begin{align}
  \sum_{i\neq j}\big(h_{ij}^2&+\tih_{ij}^2\big)=\frac12\,\sum_{i\neq
    j}\frac{y_i^2+y_j^2+6y_iy_j}{(y_i-y_j)^4}\notag\\
  &{}= 2\sum_{i\neq j}\frac{y_i^2+y_j^2}{(y_i-y_j)^4}-\frac32\,\sum_{i\neq
    j}\frac1{(y_i-y_j)^2}\notag\\
  &{}= 4\sum_{i\neq j}\frac{y_i^2}{(y_i-y_j)^4}-\frac{3N(N-1)}{8(\be+1)}\,,
    \label{hij2}
\end{align}
while from~\cite[Thm.~5.1]{AM83} it follows that
\begin{multline}
  \label{y2y4}
  720\sum_{i\neq j}\frac{y_i^2}{(y_i-y_j)^4}=\sum_iy_i^2-4(2N+\be)\sum_iy_i\\
  +2N\big(8N^2+2(4N-1)\be+3\be^2\big)\\
  -4(2 N+\be)(\be^2-2 \be-18)\sum_i\frac1{y_i}\\
  +\be(\be^3-4 \be^2-104 \be-144)\sum_i\frac1{y_i^2}\,.
\end{multline}
All the sums in the right-hand side of the latter expression have already been
evaluated in Eqs.~\eqref{yiyim1} and~\eqref{hi2}, except the first one. In
order to compute this sum, we multiply Eq.~\eqref{sum1} by $y_i$ and
sum over $i$, obtaining
\begin{align*}
  \sum_{i\neq j}\frac{2y_i^2}{y_i-y_j}&=\sum_{i\neq
    j}\frac{y_i^2-y_j^2}{y_i-y_j}=\sum_{i\neq j}(y_i+y_j)\\&
  =2(N-1)\sum_iy_i=\sum_iy_i^2-\be\sum_iy_i\,,
\end{align*}
and hence, by Eq.~\eqref{yiyim1},
\begin{equation}
  \label{yi2}
  \sum_iy_i^2=N(N+\be-1)(2N+\be-2)\,.
\end{equation}
Substituting the value of the latter sum in Eq.~\eqref{y2y4} and using~\eqref{yiyim1} and~\eqref{hi2}
we obtain
\begin{equation}\label{y2y4final}
\sum_{i\neq j}\frac{y_i^2}{(y_i-y_j)^4}=\frac{N(N-1)\big((2N+5)\be+2N+14\big)}{144(\be+1)}\,.
\end{equation}
Equation~\eqref{sigma2} now follows by inserting~\eqref{hi2}, \eqref{hijthij} and~\eqref{hij2}-\eqref{y2y4final}
into Eq.~\eqref{sigma2exp}.

\section{Computation of the minimum energy}
\label{app:minE}

In this appendix we shall obtain an explicit expression for the minimum energy $\cE^\ep_{\mathrm{min}}$
of the spin chain~\eqref{cHep}. Our starting point is Eq.~\eqref{EEE}, which implies that $\cE^\ep_{\mathrm{min}}$
is given by
\[
\cE^\ep_{\mathrm{min}}=\lim_{a\to\infty}\frac1a\,\big(E^\ep_{\mathrm{min}}-E^{\mathrm{sc}}_{\mathrm{min}}\big)
\]
in terms of the minimum energies $E^{\mathrm{sc}}_{\mathrm{min}}$ and
$E^\ep_{\mathrm{min}}$ of the scalar and spin dynamical models~\eqref{Hsc}
and~\eqref{Hep}, respectively. By the discussion in Section~\ref{sec:PF}
(cf.~Eq.~\eqref{Ens}), $\cE^\ep_{\mathrm{min}}$ is the minimum value of
$\vert\bn\vert$, where $\bn$ is any multiindex compatible with conditions
(i)--(iii) in the latter section. {}From these conditions it follows that the
multiindex $\bn$ minimizing $\vert\bn\vert$ is given by
\[
\bn=\big(\overbrace{\vphantom{1}k,\dots,k}^l\,,\overbrace{\vphantom{1}k-1,\dots,k-1}^{l_{k-1}}\,,\dots,
\overbrace{\vphantom{1}0,\dots,0}^{l_{0}}\big)\,,
\]
where $0\le l<l_k$ and $l_i$ is given by
\[
l_i=\begin{cases}
\frac m2\,, &  m \text{ even}\\[1mm]
\frac{m+\ep}2\,,\quad & m \text{ odd and } i \text{ even}\\[1mm]
\frac{m-\ep}2\,, & m \text{ and } i \text{ odd\,.}\\
\end{cases}
\]
In view of the above expression, it is convenient to treat separately the cases of even and odd $m$.
For even $m$, we have $l_i=\frac m2$ for all $i$, so that
\[
k=\lfloor 2N/m\rfloor\,, \qquad l=N \mod \frac m2\,,
\]
and
\begin{align}
  \cE^\ep_{\mathrm{min}}&=\frac m2\sum_{i=1}^{k-1}i+lk=\frac m4\,k(k-1)+lk\notag\\
  &=\frac{N^2}m-\frac N2+\frac{l(m-2l)}{2m}
  \qquad\text{($m$ even)}\,.
  \label{cEmineven}
\end{align}
Suppose now that $m$ is odd. If $k=2j$ is an even number, then
$l_0=l_2=\dots=l_{2j-2}=\frac{m+\ep}2$, $l_1=l_3=\dots=l_{2j-1}=\frac{m-\ep}2$
and thus $N=jm+l$, so that
\[
j=\lfloor N/m\rfloor\,,\qquad l=(N \mod m) <l_{2j}=\frac{m+\ep}2\,.
\]
The minimum energy in this case is thus given by
\begin{align*}
\cE^\ep_{\mathrm{min}}&=\frac{m+\ep}2\sum_{i=0}^{j-1}2i
+\frac{m-\ep}2\sum_{i=0}^{j-1}(2i+1)+2jl\\
&=mj(j-1)+j\,\frac{m-\ep}2+2jl\\
&=\frac{N^2}m-\frac{N(m+\ep)}{2m}+\frac{l(m+\ep-2l)}{2m}\,.
\end{align*}
On the other hand, if $k=2j+1$ is odd then
$l_0=l_2=\dots=l_{2j}=\frac{m+\ep}2$, $l_1=l_3=\dots=l_{2j-1}=\frac{m-\ep}2$
and thus $N=jm+l+\frac{m+\ep}2$, with $0\le l<l_{2j+1}=\frac{m-\ep}2$. Calling
$l'=l+\frac{m+\ep}2$ we have
\[
j=\lfloor N/m\rfloor\,,\qquad l'=(N \mod m)\ge\frac{m+\ep}2\,,
\]
and
\begin{align*}
\cE^\ep_{\mathrm{min}}&=\frac{m+\ep}2\sum_{i=0}^{j}2i
+\frac{m-\ep}2\sum_{i=0}^{j-1}(2i+1)\\
&\qquad+(2j+1)\Big(l'-\frac{m+\ep}2\Big)\\
&=mj(j-1)+j\,\frac{m-\ep}2+2jl'+l'-\frac{m+\ep}2\\
&=\frac{N^2}m-\frac{N(m+\ep)}{2m}+\frac{(l-m)(m+\ep-2l)}{2m}\,.
\end{align*}
Hence we can express the minimum energy for odd $m$ in a unified way as
\begin{align}
\cE^\ep_{\mathrm{min}}&=\frac{N^2}m-\frac{N(m+\ep)}{2m}
+\frac1{2m}\,\big(l-m\,\theta(2l-m-\ep)\big)\notag\\[1mm]
&\quad{\times}(m+\ep-2l)\qquad\qquad\qquad\text{($m$ odd)}\,,\label{cEminodd}
\end{align}
where
\[
l=N \mod m
\]
and $\theta(x)=1$ for $x\ge 0$ and $0$ otherwise.

\medskip

\begin{acknowledgments}
This work was partially supported by the DGI under grant no.~FIS2005-00752, and
by the Complutense University and the DGUI under grant no.~GR74/07-910556.
J.C.B. acknowledges the financial support of the Spanish Ministry of Education
and Science through an FPU scholarship.
\vspace*{.5cm}
\end{acknowledgments}
\end{document}